\documentstyle[psfig,amssymb]{mn}
\title[The Phoenix Deep Survey: X-ray properties of faint
radio sources]
{The Phoenix Deep Survey: X-ray properties of faint radio sources}

\author[A. Georgakakis et al.] {A. Georgakakis$^{1}$\thanks{email: age@astro.noa.gr},
  A. M. Hopkins$^{2}$\thanks{Hubble Fellow},  
  M. Sullivan$^3$,
  J. Afonso$^4$, I. Georgantopoulos$^1$  \\ \\
  {\LARGE B. Mobasher$^5$, L. E. Cram$^6$} \\ \\
  $^1$ Institute of Astronomy \& Astrophysics, National Observatory of
  Athens, I. Metaxa \& B. Pavlou, Penteli, 15236, Athens, Greece \\
  $^2$  Department of Physics and Astronomy, University of Pittsburgh,
  3941 O'Hara Street, Pittsburgh, PA 15260, USA \\
  $^3$ Physics Department, University of Durham,  Science Labs,
  South Road, Durham, DH1 3LE \\
  $^4$ Centro de Astronomia e Astrof\'{\i}sica da Universidade de Lisboa,
   Observat\'orio Astron\'omico de Lisboa, Tapada da Ajuda,
   1349-018 Lisboa, Portugal\\
  $5$ Space Telescope Science Institute, 3700 San Martin Drive,
 Baltimore, MD 21218, USA \\
  $6$ Australian Research Council, GPO Box 9880, Canberra ACT 2601,
  Australia \\
}

\begin{document}
\maketitle  

\begin{abstract}
In this paper we use a 50\,ks XMM-{\it Newton} pointing overlapping
with the  Phoenix Deep Survey, a homogeneous radio survey reaching
$\mu$Jy sensitivities, to explore the X-ray properties and the
evolution of star-forming galaxies.  
Multiwavelength UV, optical and near-infrared photometric data are
available for this field and are used to estimate photometric
redshifts and spectral types for all radio sources brighter than
$R=21.5$\,mag (total of 82). Faint radio galaxies with $R<21.5$\,mag
and spiral galaxy SEDs (total of 34) are then segregated into two
redshift bins with a median of $z=0.240$ (total of 19) and 0.455
(total of 15) respectively. Stacking analysis for both the 0.5--2\,keV
and 2--8\,keV  bands is performed on the two subsamples. A high
confidence level signal  ($>3.5\sigma$) is detected in the 0.5--2\,keV
band corresponding to a mean flux of $\approx 3 \times10^{-16}\, \rm
erg \, s^{-1} \, cm^{-2}$ for both subsamples. This flux translates
to mean luminosities of $\approx 5\times 10^{40}$ and  $\approx
1.5\times 10^{41}\rm \,  erg \, s^{-1}$ for the $z=0.240$ and 0.455
subsamples respectively.  Only a marginally significant signal 
($2.6\sigma$) is detected in the 2--8\,keV band for the $z=0.455$
subsample. 
This may indicate hardening of the mean X-ray
properties of sub-mJy sources at higher redshifts and/or higher
luminosities. Alternatively, this may be due to contamination of the
$z=0.455$ subsample by a small number of obscured AGNs.  
On the basis of the observed optical and X-ray properties of the faint
radio sample we argue that the stacked signal above is dominated by
star-formation with the AGN contamination being minimal.    
The mean X-ray--to--optical flux ratio and the mean X-ray luminosity
of the two subsamples are found to be higher than optically selected
spirals and similar to starbursts. We also find that the mean X-ray
and radio  luminosities of the faint radio sources studied here are
consistent with the $L_X-L_{1.4}$ correlation of local star-forming 
galaxies. 
Moreover, the X-ray emissivity of sub-mJy sources to
$z\approx0.3$ is estimated and is found to be elevated compared to
local H\,II galaxies. The observed increase is consistent with X-ray
luminosity evolution of the form $\approx(1+z)^{3}$. Assuming that our
sample is indeed dominated by star-forming galaxies this is direct
evidence for evolution of such systems at X-ray wavelengths. Using an
empirical X-ray luminosity to star-formation rate (SFR) conversion
factor we estimate a global SFR density at $z\approx0.3$ of
$0.029\pm0.007 \rm  M_{\odot}\,yr^{-1}\,Mpc^{-3}$. This is found to be
in fair agreement with previous results based on galaxy samples
selected at different  wavelengths.     
\end{abstract}


\section{Introduction}\label{sec_intro}
Deep X-ray surveys with the {\it Chandra} observatory (Brandt et
al. 2001a; Mushotzky et. al. 2000) have demonstrated beyond any doubt
the appearance of `normal' (i.e. non-AGN dominated) star-forming
galaxies at faint flux  limits, $f(\rm
0.5-2\,keV)\approx10^{-16}\rm\,erg\,s^{-1}\,cm^{-2}$ (Brandt et
al. 2001b; Hornschemeier et al. 2002a; Alexander et  al. 2002; Bauer
et al. 2002). These systems are detected in increasing numbers with 
decreasing X-ray flux and are likely to outnumber
AGNs below $f(\rm 0.5-2\,keV) \approx 10^{-17} \rm \, erg \, s^{-1}
\, cm^{-2}$ (Hornschemeier  et al. 2002b; Ranalli, Comastri \& Setti
2003). The steeply increasing X-ray source counts of
star-forming  galaxies  also suggest evolution. Our knowledge on 
this key issue, however, remains sparse.
Theoretical models predict moderate to strong X-ray evolution
for these systems at $z<1$ (Ghosh \& White 2001) but a tight
observational constraint remains to be obtained.  
Brandt et al. (2001b) studied the mean X-ray properties of bright
spirals at $z\approx0.5$ using the 500\,ks {\it Chandra} exposure of
the HDF-North. They find that the mean X-ray  luminosity of their
sample is elevated compared to local spirals suggesting evolution
consistent with the Ghosh \& White (2001) models. Differences between
the $L_B$ distribution of their sample and 
that of local spirals, however, may also be responsible for the
observed increase.  Hornschemeier et al. (2002a) used stacking
analysis to investigate the mean properties of $z\approx1$ spirals in
the {\it Chandra} Deep Field North. They found evidence for an
increase in the mean $L_X/L_B$ ratio compared to local star-forming 
systems. Statistical uncertainties, however, did not allow firm
conclusions to be drawn. 

Clearly these studies, although of key significance for understanding
the X-ray properties of star-forming galaxies at moderate to high
redshifts, cannot strongly constrain their evolution. 
To address this issue a number of observational challenges need to be
resolved. Firstly, distant starbursts are X-ray faint 
and therefore their detection even with the {\it Chandra} and the 
XMM-{\it Newton} observatories is difficult. More importantly compiling 
starburst galaxy samples with well defined selection criteria over a
wide redshift range is not trivial.    

The aim of this paper is to address these issues to provide a direct
observational constraint on the X-ray evolution of star-forming
galaxies. We combine data from an ultra-deep and homogeneous radio
(1.4\,GHz) survey (the Phoenix Deep Survey; Hopkins et al. 2003)
to the limit $S_{1.4}=80\,\mu\rm Jy$ ($5\sigma$) with a single deep
(50\,ks) XMM-{\it Newton} pointing covering an area of 30\,arcmin
diameter. Compared 
to previous studies this dataset, the Phoenix/XMM  survey, has the
advantage of deep wide area $\mu$Jy radio  observations.  Such
observations have been shown to be efficient in identifying
actively star-forming galaxies to $z\approx1$ with small
contamination by AGNs (10--20 per cent; Georgakakis et al. 1999; Barger
2002; Bauer et al. 2002; Chapman et al. 2003). 
Also, the insensitivity of radio wavelengths to dust obscuration
suggests that deep radio surveys are unique for compiling starburst
samples to $z\approx1$ with uniform selection criteria free from dust
induced biases. A minor drawback of radio selection is that it may
underestimate  the SF activity in low luminosity galaxies 
($\approx0.01L_{\star}$; Condon et al. 1992; Bell 2003; Chapman et
al. 2003). These systems are believed to substantially contribute to
the star-formation rate density in the local universe (e.g. Wilson  et
al. 2002).   
Keeping this caveat in mind, the Phoenix/XMM survey still offers an
excellent opportunity to explore the X-ray evolution of star-forming
galaxies.

Section \ref{phoenix} presents the multiwavelength data available for
the Phoenix Deep Survey, section \ref{xdata} describes the X-ray
data while section \ref{sample} details the sample used in the present
study. The stacking technique is outlined in section \ref{stacking},
and the results are presented in section \ref{results}.  Section \ref{agn}
argues against AGN contamination of the present radio selected sample
while our results are discussed in section \ref{discussion}. Finally,
section \ref{conclusions} summarises  our conclusions. For comparison
with previous studies throughout this paper we adopt $\rm
H_{o}=65\,km\,s^{-1}\,Mpc^{-1}$ and  $q_o=0.5$ but we also give
results for the $\rm \Omega_M=0.3$,  $\rm \Omega_\Lambda=0.7$ and $\rm
H_{o}=65\,km\,s^{-1}\,Mpc^{-1}$ cosmology.  

\section{The Phoenix Deep Survey}\label{phoenix}
The Phoenix Deep Survey (PDS) is an on-going program aiming to study
the nature and the evolution of sub-mJy and $\rm \mu$Jy radio 
galaxies. The radio observations were carried out at the Australia
Telescope Compact Array (ATCA) at 1.4\,GHz during several 
campaigns between 1994 and 2001 in the 6A, 6B and 6C array
configurations. The data cover a 4.56 square degree area centered
at RA(J2000)=$01^{\rm h}11^{\rm m}13^{\rm s}$
Dec.(J2000)=$-45\degr45\arcmin00\arcsec$.
A detailed description of the radio observations, data reduction and
source detection are discussed by Hopkins et al. (1998, 1999,
2003). The observational strategy adopted resulted in a radio map that 
is highly homogeneous within the central $\rm \approx1\,deg$ radius. 
Nevertheless, the $1\sigma$ rms noise increases from $\rm 12\mu Jy$ at
the most sensitive region to about $\rm 90\mu Jy$ close to the field
edge. The final catalogue consists of a total of 2058 radio sources 
to a limit  of 60\,$\rm \mu$Jy ($5\sigma$; Hopkins et al.  2003).  

Follow-up optical photometric observations of the entire PDS
in the $V$ and $R$-bands were obtained at the Anglo-Australian
Telescope (AAT) during two observing runs in  1994 and 1995
September. A detailed description of these observations including data
reduction, photometric calibration, source extraction and optical
identification are presented by Georgakakis et al. (1999). In brief,
this dataset is complete to $R=22.0$\,mag and allows optical
identification of about 50 per cent of the radio sample. An on-going
spectroscopic program aiming to obtain spectral information for the
optically identified sources is underway using the 2dF facility at the
AAT. At  present redshifts and spectral classifications are available
for over 300 sources brighter than $R=21.5$\,mag. This optical
magnitude limit is the only selection applied to the radio sample for
follow-up spectroscopy. Part of this large dataset is presented by
Georgakakis et al. (1999). Galaxies are grouped on the basis of
spectral features and diagnostic emission-line ratios into (i) systems
exhibiting absorption-line features only, (ii) star-forming galaxies,
(iii) narrow emission line Seyfert 2s, (iv) broad line Seyfert 1s and
(v) ``unclassified'' objects. The latter have at least one narrow
emission line identified in their optical spectra (allowing redshift
determination) but the poor S/N ratio, or the small number of emission
lines within the observable window, or the presence of instrumental
features contaminating emission lines prevented a reliable spectral
classification (Georgakakis et al. 1999). 

A new set of high quality deep multiwavelength $UBVRI$
photometric data has recently been obtained for a subregion of the
PDS partly overlapping with the Phoenix/XMM  field using the Wide
Field Imager at the AAT ($BVRI$-bands) and the ESO 2.2\,m ($U$-band)
telescopes complete to $I\approx24$ and
$U\approx22.5$\,mag respectively.  Near-infrared (NIR) photometric
observations ($J$ and $K$-bands) complete to $K\approx18$\,mag have
also been obtained for the 30\,arcmin diameter area of the Phoenix/XMM 
survey  using OSIRIS  at the CTIO 1.5\,m telescope. The UV, optical
and NIR  data will be presented in a series of forthcoming papers.      

\section{The X-ray data}\label{xdata}
A subregion of the PDS centered at
RA(J2000)=$01^{\rm h}12^{\rm m}52^{\rm s}$; 
Dec.(J2000)=$-45^{\circ}33^{\prime}10.0^{\prime\prime}$ was surveyed
by the XMM-{\it Newton} on 2002 May 5. The observation consists
of a single pointing with an exposure time of $\approx50$\,ks. The
EPIC (European Photon Imaging Camera; Str\"uder et al. 2001; Turner et
al. 2001) cameras were operated in full frame mode with the medium
filter applied.    

The {\it XXM-Newton} data have been analysed using the Science
Analysis Software (SAS 5.3). Event files for the PN and the two MOS
detectors have been produced using  the {\sc epchain} and {\sc
emchain} tasks of SAS respectively. The event files were screened for
high particle  background periods by rejecting times with 0.5--10\,keV
count rates higher than 20 and 6\,cts/100s for the PN and the two MOS
cameras respectively. 
The adopted count rates are a trade off between maximum effective
exposure time and low particle background contamination. Higher
thresholds do not significantly increase the exposure time while lower
thresholds severely reduce the effective exposure time. 
 The PN and MOS good time intervals are 39444 and
41273\,s respectively. The  difference between the PN and the MOS
exposure times is due to varying start and end times of the
detectors. Only events  corresponding to patterns  0--4 for the PN  
and 0--12 for two MOS cameras have been kept. To increase the
signal--to--noise ratio and to reach fainter fluxes the PN and the MOS
event files have been combined into a single event list using the {\sc
merge} task of SAS. 

Images in celestial coordinates with pixel size of 4.35\,arcsec have
been extracted in the spectral bands 0.5--8\,keV (total), 0.5--2\,keV
(soft) and 2--8\,keV (hard) for the merged event file. Events with
energies below 0.5 or above 8\,keV are not used here because of the
reduced XMM-{\it Newton} effective area at these energies. Also, below
0.5\,keV the background is elevated due to the Galactic X-ray emission
component (Lumb et al. 2002). Therefore, photons with energies $<0.5$
and $>8$\,keV primarily increase the background and do not improve the 
signal--to--noise ratio of the final image. Exposure maps accounting
for  vignetting, CCD gaps and bad pixels  have been constructed for
each spectral band.  

Source detection was independently performed in the total
(0.5--8\,keV), soft (0.5--2\,keV) and hard (2--8\,keV) band  images using
the {\sc ewavelet} task of SAS with a significance threshold of
$4\,\sigma$. A byproduct of the source  extraction algorithm is the
construction of background maps for each spectral band. We detect
137, 128 and 88 X-ray sources in the  0.5-8, 0.5--2 2--8\,keV spectral
bands respectively. The $4\,\sigma$ limiting fluxes in these spectral
bands are $f(\rm 0.5 - 8 \,keV) \approx 10^{-15}$, $f(\rm 0.5 - 2
\,keV) \approx 9 \times 10^{-16}$ and $f(\rm 2 - 8  \,keV) \approx
2 \times 10^{-15} \rm \, erg \, s^{-1} \, cm^{-2}$. A small number of
X-ray faint sources are only detected in either the soft (total of 11)
or the hard (total of 9) bands and are missed from the total band due
to the elevated background. A detailed analysis of the nature of the
X-ray sources detected in the Phoenix/XMM survey will be presented in
a forthcoming paper (Georgakakis et al. in preparation). To convert
counts to flux the Energy Conversion Factors (ECF) of individual
detectors are calculated assuming a power law spectrum with
$\Gamma=2.0$ (e.g. Bauer et al. 2002) and Galactic absorption
$N_H=2\times 10^{20} \rm {cm^{-2}}$ appropriate for the PDS. The mean
ECF for the  mosaic of all three detectors is estimated by weighting
the ECFs of individual detectors by the respective exposure time.  For
the encircled energy correction, accounting for the energy  fraction
outside the aperture within which source counts are accumulated, we
adopt the calibration 
performed by Ghizzardi (2001a, 2001b). These studies use both PN and
MOS observations of point sources to formulate the XMM-{\it Newton}
PSF for different energies and off-axis angles. In particular, a King
profile is fit to the data with parameters that are a function of both
energy and off-axis angle.  The encircled energy correction for the
merged PN+MOS image is estimated by weighting the corrections of
individual detectors by the respective exposure time. In any rate 
the difference between the PN and MOS encircled energy corrections
found by Ghizzardi (2001a, 2001b) is negligible.

\section{The radio sample}\label{sample}
A total of 204 radio sources lie within the 30\,arcmin diameter 
region covered by the Phoenix/XMM survey. This field lies within the
most homogeneously covered region of the PDS but is offset from
the most sensitive area of the radio map by about 0.30\,deg.
Therefore, the completeness limit of the  radio observations in that
30\,arcmin diameter  region is $\rm \approx80\,\mu Jy$ ($5\sigma$). A
total of 82 out of the 204 radio sources have optical  counterparts
brighter than $R=21.5$\,mag. Of the 82 sources with $R<21.5$\,mag a
subsample of 31 have spectroscopic data and secure redshifts. The
spectroscopic sample comprises 7 absorption-line systems likely to be
E/S0, 10 star-forming galaxies, 1 broad-line AGN, 1 Seyfert-2 and 12
``unclassified'' narrow emission line sources. As discussed in section
\ref{phoenix}, the latter group of galaxies have poorly constrained
optical spectral properties and their nature (AGN, star-formation)
remains uncertain. In the next section we discuss evidence suggesting
that although some  of these sources are expected to be AGNs, many
(especially at $z<0.4$) are likely to be starbursts. 

For the remaining radio sources with $R<21.5$\,mag and no
spectroscopic information (total of 61) we exploit the existing
broad-band data ($UBVRIJK$) to estimate photometric redshifts using
the {\sc hyper-z} code (Bolzonella, Miralles \& Pell\'o 2000).  The
{\sc hyper-z} program determines the photometric redshift of a given
object by fitting a set of template Spectral Energy Distributions
(SEDs) to the observed photometric data through a  standard $\chi^2$
minimisation technique. The template rest-frame SEDs used here are the
observed  mean spectra of four different galaxy types (E/S0, Sbc, Scd,
Im) from Coleman, Wu \& Weedman (1980) extended in the  UV and IR
regions  using the spectral synthesis models of Bruzual \& Charlot
(1993) with parameters selected to match the observed spectra.   

Figure 1 compares the photometric and spectroscopic redshift
estimates of the 31 objects with available spectroscopic observations.   
The agreement is good with $(z_{phot}-z_{spec})/z_{spec}
\approx0.09$. Moreover, Figure 1 shows there is a fair agreement
between the observed  spectral classification (emission, absorption
line) and the galaxy type of the best fit SED (ellipticals, spirals,
irregulars). Figure 2, plots the photometric and spectroscopic
redshift distributions of the $R<21.5$ radio sample. It is clear that
the agreement between the two distributions is good suggesting that
the photometric redshift estimates are reliable. 

We find that the majority of the sub-mJy sources are best fit
by late-type SEDs in agreement with previous studies suggesting that
the faint radio population is dominated by star-forming spirals. We
find 65 
systems with spiral/irregular type SEDs and 17 radio sources with
elliptical best fit SEDs with redshifts in the  range 0.05--2.0 and
0.05--0.6 respectively. The present radio selected sample also 
includes one broad-line AGN at $z\approx1.9$ that is assigned an irregular
galaxy SED and a photometric redshift of $\approx1.7$. Despite the
agreement, the galaxy SEDs used here are not suitable to either
estimate QSO photometric redshifts or to identify such sources within
the sample. In the absence of optical spectroscopy distant QSOs can be
identified by their unresolved optical light profile. A total of 7
radio sources have optical counterparts that are unresolved in the
$R$-band images and are excluded from the analysis as QSO candidates.  
In summary out of 82 radio sources with $R<21.5$ 65 have spiral/irregular
SEDs and 17 have E/S0 SEDs. Of the 65 sources with spiral/irregular
SEDs a total of 3 have unresolved optical light profiles and 1 is a
spectroscopically confirmed QSO. Of the 17 sources with E/S0
SEDs a total of 4 have unresolved optical light profiles.  

\begin{figure} 
\centerline{\psfig{figure=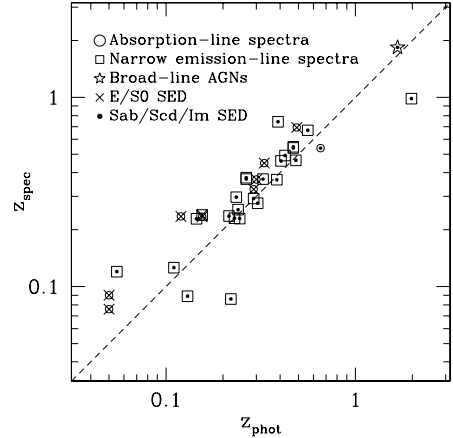,width=2.5in,height=2.5in,angle=0}} 
\caption
 {Photometric against spectroscopic redshift estimates for the
 31 sub-mJy radio sources with available spectroscopic
 observations. Open circles are radio galaxies with absorption line
 spectra, open squares correspond to systems with narrow-emission line
 spectra and stars are broad line AGNs. A cross on top of a symbol
 indicates an E/S0 best fit SED to the photometric data, while small
 dots are for Sab, Sbc or Im SEDs. There is good agreement between
 photometric and spectroscopic redshifts and between the
 spectral classification (absorption or  emission line spectra) and
 the galaxy type of the best fit SED.  
 }\label{fig_zz}    
\end{figure}

\begin{figure} 
\centerline{\psfig{figure=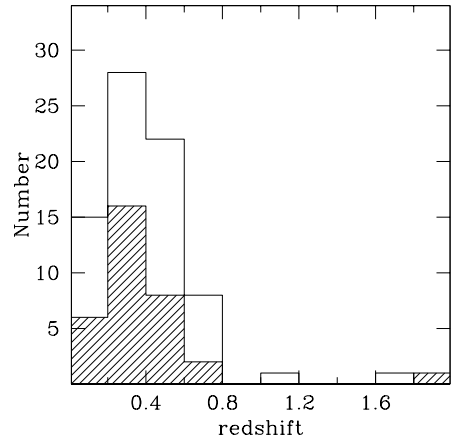,width=2.5in,height=2.5in,angle=0}} 
\caption
 {
 Redshift distributions of the sub-mJy radio sources with
 spectroscopic (hatched histogram) and photometric redshifts.
 }\label{fig_histz}    
\end{figure}

\section{Stacking procedure}\label{stacking}
Stacking methods have been extensively used in X-ray astronomy to
study  the mean properties (e.g. flux, luminosity, hardness ratios) of
well defined samples of sources that are otherwise too  faint at
X-ray wavelengths to be individually detected (e.g. Nandra et
al. 2002).    

In practice, the X-ray counts (source--plus--background) at the
position of each galaxy in the sample are added excluding 
X-ray detected galaxies. The expected background contribution is
estimated by summing the counts from regions around each
source. Assuming Poisson statistics for the counts a significance
level is estimated  for the summed signal.

To determine the size of the region within which the
source--plus--background counts are extracted we adopt the empirical 
method described by Nandra et al. (2002). Tests are performed in
which the radius of the circular aperture within which the 
source--plus--background counts are summed varies (from one trial to
the next) between 6--18\,arcsec. We find that a radius of 10\,arcsec
maximises the significance of the stacked signal and hence
this optimal extraction radius is adopted for the analysis that
follows. This extraction radius is about 2.5 times the on-axis HWHM of
the XMM-{\it Newton} PSF (Hasinger et al. 2001). To assess the
significance of the stacked signal we estimate the background from
(i) the smooth background maps  produced  by the {\sc ewavelet} task
of SAS using 20\,arcsec annuli centered on the sample galaxies and
(ii) the science images by taking the average of 60 10\,arcsec
apertures randomly positioned within annular regions centered on the
positions of the sample galaxies with inner and outer radii of 20 and
100\,arcsec respectively. In the latter method regions close to X-ray
sources ($<40$\,arcsec) are excluded  to avoid contamination from
X-ray detections in the background estimation. We find that the
significance of the stacked signal does not depend on the background
estimation method. For simplicity we use the background maps to
estimate the mean background. We note that the size of the background
aperture does not modify our results. 

To further assess the confidence level of the stacked signal we
perform extensive simulations: mock  catalogs are constructed by
randomising the positions of the radio  sample avoiding areas close to
X-ray sources 
($<40$\,arcsec). Each of the mock catalogs has the same number of
sources as the real radio catalog. We then
apply the stacking analysis to the mock catalogs and estimate 
the significance of the stacked signal in the same way as in the real
catalog. This is then repeated 10\,000 times to get the distribution
of the detection significances for the mock catalogs. This provides an
estimate of the probability of getting a statistically significant
stacked signal by chance. The detection confidence levels estimated
from the simulations are in excellent agreement with those based on
Poisson statistics  

In practice we have excluded from the stacking analysis galaxies that
are identified with X-ray sources detected by the {\sc ewavelet} 
task to the $4\,\sigma$ significance level. Such a low detection
threshold is essential to study the mean properties of X-ray weak
sources, the signal of which would otherwise be diluted by brighter 
ones. 
We find 9 radio sources with X-ray counterparts within 8\,arcsec off
the radio position: 4 of them have narrow emission-line spectra, while
the rest have no spectroscopic information but their broad band  
colours suggest spiral galaxy SEDs. Most of these X-ray detected radio
sources have hard X-ray spectra and X-ray luminosities of
$\approx10^{43}\,\rm\,erg\,s^{-1}$ suggesting AGN activity 
rather than star-formation. A detailed study of these radio sources
will be presented in a forthcoming paper.  

Moreover, counts associated with the wings of the PSF of bright X-ray
sources might erroneously increase the stacking signal
significance. Therefore, to avoid contamination of the stacking signal  
from nearby X-ray sources we have excluded from the analysis a total
of 12 radio galaxies that lie close (i.e. 8--40\,arcsec) to X-ray
detections. A total of 21 radio sources, either 
lying close to (12 objects; 8--40\,arcsec) or being associated with
(9 objects; $<8$\,arcsec) X-ray detections, have thus been excluded.

The aim of the present paper is to explore the mean properties of
star-forming sub-mJy radio galaxies. Our sample is restricted to
sub-mJy sources with $S_{1.4}>80\,\mu$Jy (the flux density limit of
the radio catalogue) and spiral galaxy SEDs that are not associated
with or do not lie close to X-ray sources. Further, in the analysis that
follows we use only radio sources with redshift $z<0.8$, since at higher
redshifts the present sample with $R<21.5$ becomes highly
incomplete. We also exclude distant AGNs as explained above.
The final sample used in the stacking analysis comprises
34 radio galaxies that fulfill the above criteria.

\section{Results}\label{results}

\begin{table*} 
\footnotesize 
\begin{center} 
\begin{tabular}{cccc c cccc cccc cc} 
\hline 
redshift
& median
& source 
& \multicolumn{2}{c}{S+B$^{a}$}  
& \multicolumn{2}{c}{B$^{b}$} 
& \multicolumn{2}{c}{SNR$^{c}$} 
& \multicolumn{2}{c}{$<f_X>^{d}$}
& \multicolumn{2}{c}{$<L_X>^{e}$} 
& $L_B^{f}$
& $L_{1.4}^{g}$ \\

range
&redshift 
& number
& SB & HB
& SB & HB
& SB & HB
& SB & HB
& SB & HB
&  
&  \\
0.0--0.8 & 0.291 
& 34 
& 467 & 380 
& 365.3 & 345.1
& 5.3 &  1.9
& $2.9\pm1.0$ &  $<10.3$
& $6.8\pm2.3$ &  $<24.5$
& $1.39$& $5.0$\\

0.0--0.3 & 0.240 
& 19 
& 267 & 204
& 208.1 & 199.7
& 4.1 &  0.3
& $3.0\pm1.4$ &  $<9.5$
& $4.8\pm2.2$ &  $<15.2$
& $1.38$& $2.0$\\

0.3--0.8 & 0.445
& 15 
& 200 &  176 
&  157.2 & 145.3
& 3.4 & 2.6
& $2.7\pm1.4$ & $<17.0$
& $15.4\pm7.8$ & $<98.6$
& $1.40$& $6.3$\\

%
0.0--0.3$^h$ & 0.177
& 6
& 86 & 52
& 58.89 & 55.7
& 3.5 & -0.5
& $5.9\pm2.9$ & $<16.9$
& $5.0\pm2.5$ & $<14.4$
& $1.61$& $1.0$\\

\hline
\multicolumn{14}{l}{$^a$Source+Background counts in the soft (SB) and
the hard (HB) bands} \\ 
\multicolumn{14}{l}{$^b$Background counts in the soft (SB) and the hard
(HB) bands} \\  
\multicolumn{14}{l}{$^c$Significance of detection in background standard
deviations in the soft (SB) and the hard  (HB) bands} \\   
\multicolumn{14}{l}{$^d$X-ray flux in units of $\rm
10^{-16}\,erg\,s^{-1}\,cm^{-2}$ in the soft (SB) and the hard
 (HB) bands} \\
\multicolumn{14}{l}{$^e$X-ray luminosity in units of $\rm
10^{40}\,erg\,s^{-1}$ in the soft (SB) and the hard
(HB) bands} \\
\multicolumn{14}{l}{$^f$B-band luminosity in units of $\rm
10^{43}\,erg\,s^{-1}$} \\
\multicolumn{14}{l}{$^g$1.4\,GHz luminosity density in units of $\rm
10^{22}\,W\,Hz^{-1}$} \\
\multicolumn{14}{l}{$^h$Spectroscopically confirmed star-forming
galaxies on the basis of diagnostic line ratios} \\

\end{tabular} 
\end{center} 
\caption{ Stacking analysis results in the soft (0.5--2\,keV) and hard
(2--8\,keV) bands for the different samples of the Phoenix/XMM faint radio
galaxies.}\label{stacking_tbl}  
\normalsize  
\end{table*} 

The sub-mJy radio sources used in the stacking analysis are split
into two independent redshift bins $z=0.0-0.3$ and $z=0.3-0.8$ with median
redshifts of 0.240 and 0.445 respectively, selected to have similar
numbers of sources. The stacking analysis has also been performed on the
subsample of radio sources with available optical spectroscopy and
diagnostic line ratios typical of star-formation activity. For this
subsample we only consider star-forming sources with $z<0.3$. At this
redshift range the $\rm H\alpha$ emission-line lies within the
observable window of the spectroscopic observations allowing reliable
classification on the basis of diagnostic emission-line ratios. The
stacking analysis of this subsample will allow us to investigate
whether Seyfert-2s that remain unidentified due to the absence of
optical  spectroscopy for many radio sources can significantly bias
our conclusions. 

The stacking results for the above subsamples in both the soft and
the hard bands  are presented in Table \ref{stacking_tbl}. The X-ray
flux in this table is estimated  from the raw net counts after
correcting individual galaxies for (i) the effect of vignetting
estimated from the corresponding exposure map  and (ii) the energy
fraction outside the adopted aperture of 10\,arcsec using the
encircled energy corrections recently derived by Ghizzardi  (2001a,
2001b) for both the PN and the MOS detectors (see section
\ref{xdata}). The counts--to--flux conversion factor has been derived
assuming a power-law with spectral index $\Gamma=2.0$ and   Galactic
absorption (see section \ref{xdata}). The X-ray luminosities in Table
\ref{stacking_tbl} are estimated using the X-ray flux derived from the
stacking 
analysis and the median redshift of each  subsample also listed in
this table. The k-correction is estimated for a power-law with
spectral index $\Gamma=2.0$. Both the X-ray flux and luminosity are
corrected for Galactic photoelectric absorption ($N_H=2\times 10^{20}\,
\rm cm^{-2}$) but not for intrinsic absorption. 
The $L_B$  and  $L_{1.4}$ in Table \ref{stacking_tbl} are the 
medians of the rest frame luminosities of individual galaxies. 
The $B$-band k-correction for a given radio source is derived
from the  best fit SED estimated by {\sc hyper-z}. 
At 1.4\,GHz we assume a power law SED  
of the form $f_{\nu}\sim\nu^{-0.8}$ to calculate the k-correction and
to estimate the rest frame  $L_{1.4}$ of individual radio sources.

In the soft-band a statistically significant signal is obtained for
all the subsamples in Table 1. Stacking of the hard band counts,
gives a marginally significant signal for the high redshift subsample
at the $2.6\sigma$ confidence level. For the whole sample the hard band
stacked signal is significant at the $1.9\sigma$ level but this is
primarily due to high-$z$ rather than low-$z$ sources. The
marginally significant hard band signal of the $z=0.3-0.8$ sub-sample may
be due to contamination by few AGNs that lie just below the detection
threshold. This is further discussed below. Alternatively this may
suggest hardening of the mean X-ray properties of radio sources at
higher redshifts and/or at higher radio powers. It is
thus possible that the present high-$z$ radio subsample has X-ray
spectral properties different to those of the low-$z$ one. Deeper
X-ray data are required to provide stronger constraints on the X-ray
spectral properties of the $z=0.3-0.8$ radio sources.   


For the 2--8\,keV  band $3\sigma$ upper limits are estimated  
for the stacked signal assuming Poisson statistics. Also,
spectroscopically confirmed star-forming radio sources exhibit
a statistically significant stacked signal and have 0.5--2\,keV X-ray
flux and luminosity similar to that obtained for the $z=0.0-0.3$
subsample spanning a similar redshift range.  This suggests that
powerful AGNs do not significantly bias our results at least for the
$z=0.0-0.3$ redshift bin.


Comparison of the estimated soft-band count rates with the hard-band
upper limits can constrain the spectral properties of the sub-mJy
radio  sources with spiral galaxy SEDs. For the $z=0.0-0.3$ redshift bin
the observed count rates at the median redshift $z=0.240$ translate to
an upper limit of $N_{H}\approx4\times 10^{21}\rm \,cm^{-2}$ assuming
a power law with spectral index $\Gamma=1.7$. This increases
to $N_{H}\approx8\times 10^{21} \rm \,cm^{-2}$ for
$\Gamma=2$. By constraining the column density to the Galactic value of
$N_{H}=2\times 10^{20}\rm\,cm^{-2}$ a lower limit to the
spectral index $\Gamma>1.3$ is inferred. These 
constraints, although not very tight, are consistent with the spectral
properties of nearby star-forming spirals (e.g. Fabbiano 1989). For
the $z=0.3-0.8$ subsample, the  observed count rates  translate to an
upper limit of $N_{H}\approx1\times 10^{22}\rm \,cm^{-2}$ assuming a
spectral index $\Gamma=1.7$. This  will increase to $N_{H}=2\times  
10^{22} \rm \,cm^{-2}$ for $\Gamma=2$. Keeping the column density
constant at the Galactic value $N_{H}\approx2\times
10^{20}\rm\,cm^{-2}$ we find a lower limit to the spectral index
$\Gamma>0.8$. If the marginally significant hard band count rate is
taken at face value, however, we estimate a hydrogen column density of
$N_{H}\approx1\times10^{22}\rm\,cm^{-2}$ ($\Gamma\approx2.0$). This
column density may indicate either low-luminosity AGN (LLAGN) activity
or heavily obscured AGN/star-formation for the  $z=0.3-0.8$
subsample. Using the empirical relation between X-ray absorption and
optical extinction of Gorenstein (1975) the above column density
translates to $E(B-V)\approx1.4$. Leech et al. (1988) obtained optical
spectra for high Galactic latitude IRAS  sources (both starbursts and
AGNs) and found a mean $E(B-V)\approx1.3$ comparable to the reddening
estimated  here. Veilleux et al. (1995) used high quality nuclear
optical spectra of luminous IRAS sources. For the sub-sample
classified H\,II galaxies they estimate a mean $E(B-V)\approx1$ which
is somewhat lower than the value estimated here.

We also investigate the possibility that the observed stacked signal
is primarily due to a few bright X-ray sources below the detection
threshold. This is important especially if some of these sources are
AGNs and thus not representative of the star-forming galaxy
population studied in this paper. We explore this issue by assigning 
detection significances to individual radio sources and then excluding
sources above a given threshold. The detection significance is
calculated by the ratio of net source counts to the square root of the
background estimated as described in section \ref{stacking}.  

For the whole ($z=0.0-0.8$) sample a significant 0.5--2\,keV signal is
still obtained after removing sources above $3\sigma$ (total of 3).
These sources have relatively low X-ray--to--optical luminosity ratios
$\log L_X / L_B \approx-2$, typical of starbursts. Clearly,
deeper X-ray observations are required to further address this issue.

For the 2--8\,keV band all sources have  hard band detection
significances below the $3\sigma$ level. Removing the 3 higher
significance sources ($>1.7\sigma$) reduces the confidence level 
of the hard band stacked signal to $\approx1\sigma$
for  both the $z=0.0-0.8$   and $z=0.3-0.8$ subsamples
respectively. Therefore, it is possible that contamination by few
obscured AGNs is responsible for the apparent hardening of the mean
X-ray spectral properties of the high-$z$ sub-sample. 

We note that for flat $\rm \Lambda$ cosmology ($\rm \Omega_{M}=0.3$,
$\rm \Omega_{\Lambda}=0.7$) the luminosities estimated in Table 
\ref{stacking_tbl}  will increase by $\approx 24$ and $\approx 40$
per cent for the  $z=0.0-0.3$ and  $z=0.3-0.8$ subsamples respectively

\begin{figure} 
\centerline{\psfig{figure=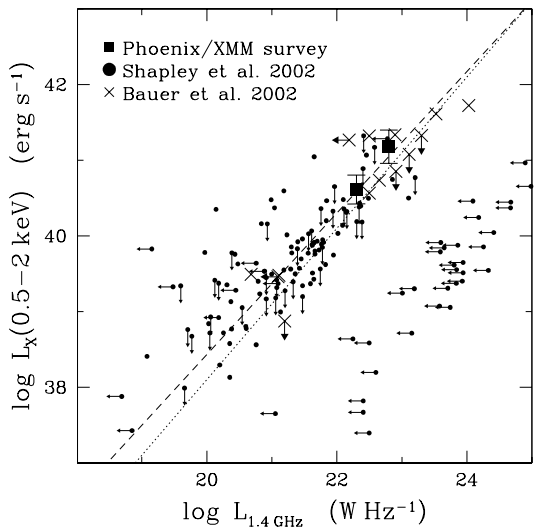,width=3in,height=3in,angle=0}} 
\caption
 {
 Soft band 0.5--2\,keV X-ray luminosity against 1.4\,GHz radio
 power. The stacking analysis results for the PDS
 radio sources in the redshift range $z=0.05-0.3$ and $0.3-0.8$ are shown
 with the filled squares. The small filled circles are local galaxies
 from the  sample compiled by Shapley et al. (2001). The crosses are
 star-forming faint radio sources from the sample of Bauer et
 al. (2002). The dashed line shows the  best fit $L_X-L_{1.4}$
 relation derived by Bauer et al. (2002) shifted to the 0.5--2\,keV
 band as  described in the text. The dotted line is the best fit
 $L_X-L_{1.4}$ relation derived by Ranalli et al. (2003). 
 }\label{fig_lxl14}    
\end{figure}

\begin{figure} 
\centerline{\psfig{figure=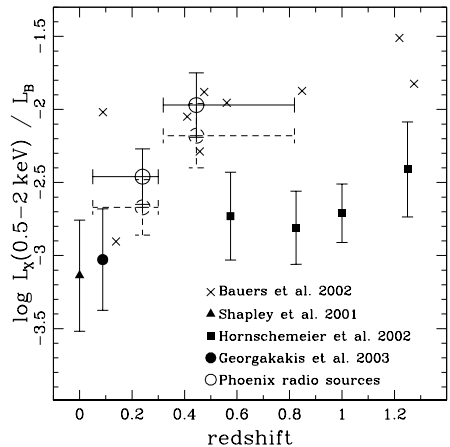,width=2.5in,height=2.5in,angle=0}} 
\caption
 {
 $\log L_X({\rm 0.5-2\,keV}) / L_B $ against redshift. Open circles
 are the stacking analysis results for the late type (Sbc, Scd and Im)
 PDS radio sources in the redshift range $z=0.05-0.3$ and $0.3-0.8$. The
 points are plotted at the median $z$, while the horizontal
 errorbars indicate the extent of the redshift bins. The dashed line
 symbols represent our stacking analysis  $\log L_X / L_B $ estimates
 after correcting for the difference in $L_B$ between optical and
 radio selected samples (see text for details). Also shown
 are the mean  $\log L_X / L_B $ of local spirals (triangles; Shapley
 et al. 2001), $z\approx0.1$ spirals (filled circle; Georgakakis et
 al. 2003) and the stacking analysis results  for distant late type
 galaxies in the {\it Chandra} Deep Field North (squares;
 Hornschemeier  2002a). The crosses are the  $\log L_X / L_B $ ratio
 for narrow emission line radio galaxies in the {\it Chandra} Deep
 Field North from the sample
 of Bauer et al. (2002).
 }\label{fig_lxlb}    
\end{figure}

\section{AGN contamination}\label{agn}

The aim of the present paper is to explore the mean properties of
star-forming sub-mJy radio galaxies and we have attempted
to exclude sources that show evidence for strong AGN activity
from the sample used in the stacking analysis. We argue that the AGN
contamination of the present radio selected sample is small:

\begin{itemize}

\item Radio selection at $\mu$Jy flux densities has been shown to be
efficient in finding actively star-forming systems to
$z\approx1$. Only $\approx10-20$ per cent of the  faint radio
population are believed to be AGNs (Georgakakis et al. 1999; Barger et 
al. 2002; Bauer et al. 2002; Chapman et al. 2003).

\item A total of 7 radio sources with optical counterparts exhibiting
unresolved light profiles in the $R$-band, likely to be distant QSOs,
have been identified and excluded from the sample.       

\item Within the spectroscopic sample only 5 out of the 31 radio
sources with spectroscopic observations show evidence for  
Seyfert-1 or 2 type activity. One of these sources is a distant
($z\approx1.9$) QSO, another source is classified Seyfert-2 on the
basis of diagnostic optical emission line ratios and the remaining 
three belong to the group of spectroscopically ``unclassified''
sources  (see section \ref{phoenix}). These three sources have X-ray
counterparts with X-ray spectral properties indicative of AGN
activity (e.g. hard X-ray spectra, X-ray luminosities in excess of
$\rm \approx10^{42}\,erg\,s^{-1}$). All of these sources have been
excluded from the stacking analysis. 

\item For the group of spectroscopically ``unclassified'' radio
sources (see section \ref{phoenix}) {\it without} X-ray counterparts
to the sensitivity limit of the existing observations (total of 9) we 
can constrain their nature by estimating upper limits to their
$L_X/L_B$ and $L_X$.  For sources with $z<0.4$ (4 out of 9) we find
$\log f_X/f_{opt}<-2$  and $L_X(\rm 0.5-2
\,keV)<2\times10^{41}\,\rm\,erg\,s^{-1}$ consistent with star-formation  
rather than AGN activity. For sources with $z>0.4$ (5 out of 9) the
estimated upper limits are not as stringent: most of these sources
typically have $\log f_X/f_{opt}<-1.2$  and $L_X(\rm 0.5-2
\,keV)<5\times 10^{41}\,\rm\,erg\,s^{-1}$. These limits  exclude the
possibility of powerful AGNs and suggest starbursts, obscured AGNs or
LLAGNs. In the case of obscured AGNs the soft 0.5--2\,keV band X-ray
emission from the AGN is expected to be heavily obscured. The X-ray 
emission in the 0.5--2\,keV spectral band is thus likely to be dominated by
stellar processes (e.g. NGC\,6240; Vignati et al. 1999).
Deeper X-ray  observations are required to elucidate the detailed nature of
these sources. Nevertheless, the  $z<0.4$ sub-sample for which our 
observations can put tight constraints is dominated by
star-formation. 

\item In Table \ref{stacking_tbl} we estimate mean
X-ray--to--optical luminosity ratios of $\log L_X / L_B \approx-2.4$ 
and --2.0 for the 0.240 and 0.445 redshift subsamples
respectively. These ratios although elevated compared to those found
for optically selected spiral galaxy samples (Hornschemeier et
al. 2002a; Georgakakis et al. 2003) are typical of actively
star-forming galaxies rather than AGN dominated systems (Alexander et
al. 2002). Moreover, the 0.5--2\,keV X-ray luminosities in the two
redshift bins are $\approx5\times 10^{40}$  and $\approx1.5\times
10^{41}\rm \,erg\,s^{-1}$ respectively. These luminosities are
intermediate to Milky Way type spirals ($\approx10^{40}\rm
\,erg\,s^{-1}$; Warwick et al. 2002)  and luminous starbursts like
NGC\,3256 ($\approx10^{42}\rm \,erg\,s^{-1}$; Moran, Lehnert \&
Helfand 1999).   

\item The estimated mean X-ray flux ($f_X(\rm
0.5-2\,keV)=5.9\times10^{-16}\rm\,erg\,s^{-1}\,cm^{-2}$) and
luminosity ($L_X(\rm 0.5-2\,keV) = 5.0 \times 10^{40} \rm \, erg \,
s^{-1}$) of spectroscopically confirmed $z<0.3$ star-forming radio   
sources (see Table  \ref{stacking_tbl})  are similar to those obtained
for the {\it all} $z<0.3$ radio sources with spiral SEDs ($f_X(\rm
0.5-2\,keV)=3.1\times10^{-16}\rm\,erg\,s^{-1}\,cm^{-2}$; $L_X(\rm
0.5-2\,keV) = 4.8 \times 10^{40} \rm \, erg \, s^{-1}$) . Hence, any  
contribution from AGNs/QSOs at least for the $z=0.0-0.3$ subsample is
minimal. 


\item The mean X-ray and radio luminosities of the faint radio sources
studied here are consistent with the $L_X-L_{1.4}$ correlation of
local star-forming galaxies. This is discussed in detail in the
next section. 

\end{itemize}

The evidence above suggest that the detected stacked X-ray signal is
consistent with star-formation activity. However,  despite our effort to
minimise the AGN contamination we cannot exclude the possibility of a
small fraction of low-luminosity or obscured AGNs within our radio
selected sample.   

\section{Discussion}\label{discussion}
\subsection{X-ray properties of the faint radio population}
The sub-mJy and $\rm \mu$Jy  radio population is believed to
comprise a large fraction of starburst galaxies out to $z\approx1$
(Benn et al. 1993; Hopkins et al. 1998; Georgakakis et al. 1999;
Chapman et al. 2003). The enhanced SFR in these systems results in
increased radio emission due to both  thermal emission and synchrotron
radiation in supernovae remnants. Moreover, studies of local
star-forming and spiral galaxies demonstrate that these systems are
also powerful X-ray emitters (Fabbiano 1989; Read, Ponman \&
Strickland 1997; Dahlem, Weaver \& Heckman 1998; Read \& Ponman 2001)
with luminosities that in some cases exceed $\approx10^{42}\,\rm
erg\,s^{-1}$  (e.g. NGC\,3265;  Moran et al. 1999). The X-ray emission
in these galaxies originates from a combination of evolved stellar
point sources (X-ray binaries) and diffuse hot ($T \approx 1-8\times
10^{6}$\,K) gas heated by supernovae shocks (Read, Ponman \&
Strickland 1997; Read \& Ponman 2001). Since the physical processes
giving rise to both the X-ray and the radio emission have the same origin,
i.e. enhanced SFR, a correlation should be expected between the X-ray
and radio luminosities of star-forming galaxies and sub-mJy radio
sources.   


Shapley, Fabbiano \& Eskridge (2001) compiled a large sample of local
spirals with X-ray data from the {\it Einstein} observatory and
multiwavelength observations. They find a highly significant
correlation between the X-ray and radio luminosities suggesting an
association between X-ray emission and star-formation
activity. However, although Shapley et al. (2001) exclude AGN
dominated systems from the analysis, their sample may be contaminated
by LLAGNs.  Ranalli et al. (2003) used the atlas of
optical nuclear spectra of Ho et al. (1997) to compile a carefully
selected sample of  nearby star-forming galaxies with available X-ray
and radio data. The advantage of their sample is that the high quality
nuclear spectra of Ho et al. (1997) provide reliable spectral
classification of the central source (star-formation/AGN) thus,
minimising the contamination from LLAGN. Moreover, use of {\it
ASCA} and {\it BeppoSAX} X-ray data allow  Ranalli et al. (2003) to
extend their study of star-forming galaxies to hard X-rays
(2--10\,keV).  These authors report a tight correlation between radio
power (1.4\,GHz)  and X-ray luminosity in both the 0.5--2 and 
the 2--10\,keV spectral bands. 

Bauer et al. (2002) investigated the association between faint X-ray
and radio source populations detected in the 1\,Ms {\it Chandra} Deep
Field North (CDF-N). They found that the majority of X-ray sources
with narrow emission-line spectra also have $\mu$Jy radio counterparts
and are likely to be dominated by star-formation activity. Also,
for the star-forming galaxy subsample they find an $L_X-L_{1.4}$
relation similar to that obtained by Shapley et al. (2001) for local
non-AGN spirals. They argue that the evidence above suggests that the
local $L_X-L_{1.4}$ can be extended to intermediate and high redshifts
and that the X-ray emission of narrow emission line X-ray sources is
associated with star-formation activity.

Figure \ref{fig_lxl14} plots 1.4\,GHz radio power against 0.5--2\,keV
X-ray  luminosity for both the Shapley et al. (2001) local spirals and
the Bauer et al. (2002) distant sub-mJy radio sources with narrow
emission line spectra. To transform the X-ray luminosities of the
above samples to the 0.5--2\,keV band used in the present study we
assume a power-law model with a spectral index $\Gamma=2.0$. At radio
wavelengths a power-law spectral energy distribution of the form 
$f_\nu\sim\nu^{-0.8}$ appropriate for star-forming galaxies is adopted
to convert the 4.85\,GHz radio flux density of Shapley et al. (2001)
to 1.4\,GHz.  
Also shown in Figure \ref{fig_lxl14} are the stacking analysis
results of the  PDS sub-mJy radio sources in the $z=0.0-0.3$ 
and $z=0.3-0.8$ redshift bins. It is clear that there is excellent
agreement between the mean X-ray and 1.4\,GHz luminosities derived 
here and the $L_X-L_{1.4}$ relation of Bauer et al. (2002) and Ranalli
et al. (2003) for star-forming galaxies. This suggests that the
present sample of sub-mJy radio sources to $z\approx0.5$ is dominated
by starburst rather than  AGN activity. Indeed, AGN dominated systems
as well as LLAGNs have X-ray--to--radio luminosity ratios
elevated compared to those of star-forming galaxies
(Brinkmann et al. 2000; Alexander et al. 2002; Ranalli et al. 2003).  

To further investigate the nature of the faint radio sources we use
their X-ray--to--optical luminosity ratios. Figure \ref{fig_lxlb}
plots the mean $L_X/L_B$ ratio of the present sample against redshift 
in comparison with optically selected `normal' spirals. It is clear
that faint radio sources have elevated $L_X/L_B$ ratios by a factor
of 2--10 compared to optically selected spiral galaxy samples. This
is partly due to the enhanced mean $L_B$ of the present
sample in combination with the non-linear correlation between optical
and X-ray luminosities  $L_X\approx L_B^{1.5}$ reported by 
Shapley et al. (2001). 
Accounting for the differences in the mean $L_B$ of radio and 
optically selected samples will decrease our $L_X/L_B$ estimates by
about 0.2 in log space. The corrected  $L_X/L_B$ ratios are still
somewhat elevated
compared to those of Georgakakis et al. (2003) and Hornschemeier  
et al. (2002a) samples. This  may suggest elevated star-formation
activity in the radio selected sample. Nevertheless, the
apparent difference is not statistically significant. Indeed, within the
errors the $L_X/L_B$ ratio is roughly constant with redshift
suggesting that the $L_X$ of star-forming galaxies evolves with 
redshift  in the same rate as $L_B$.    

Moreover, the difference in the $L_X/L_B$ between low and high-$z$
radio selected sub-samples, although not very significant, may suggest
enhanced galaxy activity with increasing redshift or radio power. Also
shown in this Figure \ref{fig_lxlb} are the $L_X/L_B$ ratios of
individual CDF-N narrow-emission line radio sources with X-ray
counterparts (Bauer et al. 2002). We use the $I$-band magnitudes given
by Bauer et al. (2002) to estimate rest-frame $B$-band luminosity
using the SED of Sbc type galaxies. There is good agreement between
the $L_X/L_B$ of the CDF-N X-ray detected radio sources and our
stacking analysis results.     

Finally, the analysis above suggests that sub-mJy spiral galaxies
have X-ray properties similar to those of X-ray detected narrow
emission-line galaxies. Indeed, many of these sources also appear to
reside in disc galaxies with blue optical colours and enhanced X-ray
emission relative to the optical (McHardy et al. 1998), while their
optical spectra and X-ray properties suggest a mix of star-forming
galaxies and obscured AGNs. Therefore, sub-mJy galaxies and at least a
fraction of the X-ray selected narrow emission-line galaxies are
drawn from the same parent population. This is in agreement with
Bauer et al. (2002) who find a large overlap between the faint X-ray
and radio selected populations especially for the sources with narrow
emission line spectra. They argue that these galaxies are most likely
normal and starburst spirals in the redshift range $z\approx0.3-1.3$.

\begin{figure} 
\centerline{\psfig{figure=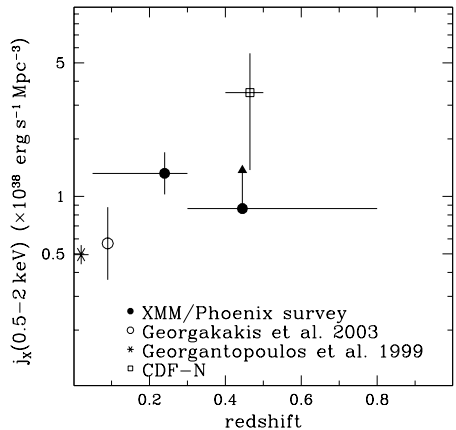,width=2.5in,height=2.5in,angle=0}} 
\caption
 {
 0.5--2\,keV X-ray emissivity, $j_X$, against redshift. Filled circles are the
 sub-mJy star-forming radio sources from the Phoenix/XMM survey in the
 redshift bins $z=0.05-0.3$ and $0.3-0.8$. These points are plotted at the
 median $z$, while the horizontal errorbars indicate the extent
 of each bin. The open   
 circle represents optically selected spirals at $z\approx0.1$ from
 the 2dF Galaxy Redshift Survey (Georgakakis et al. 2003).  The star
 corresponds to local star-forming systems (Georgantopoulos et
 al. 1999). The open square is the $j_X$ estimate for narrow emission
 line radio sources with X-ray counterparts detected in the CDF-N (see
 text for details).    
 Although the uncertainties are large, there is evidence
 for evolution of $j_X$ to $z\approx0.3$ of the form $(1+z)^{3}$. At
 higher redshifts the Phoenix/XMM survey is affected by incompleteness
 allowing only a lower limit for  $j_X$ to be estimated. The CDF-N  
 point at $z\approx0.5$ provides some evidence for  evolution at
 $z>0.3$ but the large uncertainty (due to small number statistics) do
 not allow firm conclusions to be drawn.  
 }\label{fig_jx}    
\end{figure}

\begin{figure} 
\centerline{\psfig{figure=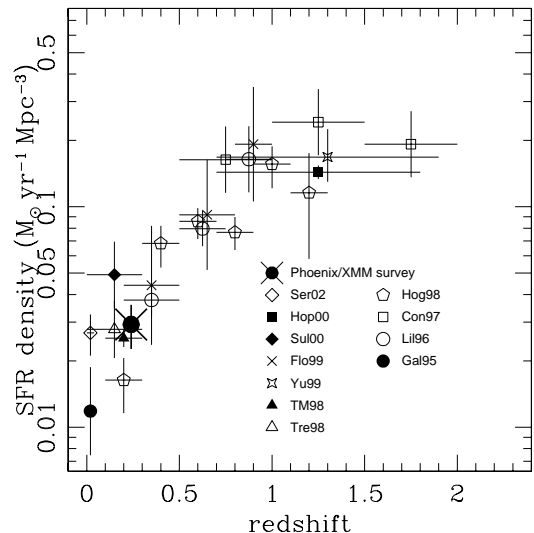,width=3in,height=3in,angle=0}} 
\caption
 {
 Global star-formation density against redshift estimated from galaxy
 samples selected at various wavelengths. SFR dependent reddening
 corrections have been applied to the data as described in Hopkins et
 al. (2001). References in this diagram are as follows: Ser02:
 Serjeant et al. 2002; Hop00: Hopkins et al. 2000; Sul00: Sullivan et
 al. 2000; Flo99: Flores et al. 1999; Yan99: Yan et al. 1999; TM98:
 Tresse \& Maddox (1998); Tre98: Treyer et al. 1998; Hog98: Hogg et
 al. 1998; Con97: Connolly et al. (1997); Lil96: Lilly et al. 1996;
 Gal95: Gallego et al. (1995). The SFR density estimated in the
 present study using stacking analysis is shown with the crossed
 filled circle. No absorption corrections have been applied. Given the
 uncertainty in converting X-ray luminosities to star-formation rates
 the agreement of our estimate at  $z\approx0.3$ with previous studies
 is encouraging.  
 }\label{fig_sfr}    
\end{figure}

\subsection{X-ray emissivity of sub-mJy galaxies}
The emissivity of galaxy samples selected at various wavelengths (UV,
optical, radio) over a range of redshifts is one of the most powerful
and widely used evolutionary tests (e.g. Lilly et al. 1996; Connolly et
al. 1997).  

The X-ray emissivity, $j_X$, of star-forming systems  has been
estimated locally by Georgantopoulos, Basilakos \& Plionis (1999).
They convolved the local optical luminosity function of the Ho et
al. (1995) sample with the corresponding $L_X-L_B$ relation based 
on {\it Einstein} data. The Ho et al. sample has the advantage of high
S/N nuclear optical spectra and hence, reliable spectral
classifications (Ho et  al. 1997) necessary to study the properties of
star-forming systems. Recently, Georgakakis et al. (2003)
compiled a sample of `normal' spirals from the  2dF Galaxy Redshift
Survey and used stacking analysis to estimate their $j_X$ at
$z\approx0.1$. The Phoenix/XMM survey provides the opportunity to
extend these results to higher redshifts, assuming that the
present sample of sub-mJy sources is indeed dominated by
star-formation activity.

For the $j_X$ determination, we first  need to estimate the effective
cosmological volume probed by the present radio selected sample
corrected for optical and radio selection biases (e.g. $R<21.5$\,mag,
$S_{1.4}>80\rm\,\mu Jy$) using the standard $1/V_{max}$ formalism 
(e.g. Lilly et al. 1996; Mobasher et al. 1999). We also account for
radio catalogue incompleteness due to (i) the varying sensitivity of
the radio observations across the field (weighting correction) and
(ii) extended low surface brightness radio sources with total flux
density above the survey limit that can be missed by the detection
algorithm (resolution correction).  These biases as well as
methods to correct for them are fully described by Hopkins et al.
(1998). We note that the rms noise level of the radio observations is
relatively uniform over the area of the Phoenix/XMM field and the resulting
weighting correction is small. Finally in estimating $j_X$ we also assume that
the sub-mJy radio sources have constant $L_X/L_{1.4}$ ratio as
indicated in Figure  \ref{fig_lxl14}. 

We estimate soft-band 0.5--2\,keV emissivities of
$(1.3\pm0.3)\times 10^{38}$ and $(0.9\pm0.1)\times 10^{38} {\rm
erg\,s^{-1}\,Mpc^{-3}}$ for the late-type sub-mJy radio sources
in the redshift bins $z=0.0-0.3$ and $z=0.3-0.8$ respectively.  
We note that the $j_X$ estimate of the  $z=0.3-0.8$ subsample is a lower
limit. This is mainly due to  the fraction of sub-mJy radio sources in
the $z=0.3-0.8$ redshift range and optical magnitudes fainter than
$R=21.5$. Future optical observations reaching fainter magnitude
limits will help address this issue. Moreover, radio sources at a
given redshift with radio luminosity fainter than the survey limit
are missed in the  $j_X$ calculation. We estimate the magnitude of
this effect using the evolving radio luminosity function of
Rowan-Robinson et al. (1993) that has been shown to reproduce the
observed radio source counts below 1\,mJy (Hopkins et al. 1998).
We find  this bias to be small for the $z=0.0-0.3$  redshift bin:
correcting for the radio sources fainter than the survey limit
increases $j_X$ by less than about 10 per cent. However, for the
$z=0.3-0.8$ bin this effect is larger, $\approx30$ per cent.   
For flat $\rm \Lambda$ cosmology ($\rm \Omega_{M}=0.3$, $\rm
\Omega_{\Lambda}=0.7$) the $j_X$ estimated here will 
decrease by $\approx10$ and $\approx30$ per cent for the  $z=0.0-0.3$ and    
$z=0.3-0.8$ subsamples respectively

Figure  \ref{fig_jx} plots the X-ray emissivity of the sub-mJy radio
sources against redshift in comparison with that of
local star-forming galaxies (Georgantopoulos et al. 1999; scaled to
the 0.5--2\,keV  band) and  $z\approx0.1$ 2dF Galaxy Redshift Survey
spirals (Georgakakis et al. 2003). 
It is clear that the $j_X$
in  Figure  \ref{fig_jx} increases from the local Universe to
$z\approx0.3$, suggesting evolution. At higher redshifts, as already
mentioned, the present study cannot further constrain the $j_X$
evolution, although with deeper optical observations of the Phoenix/XMM
field this issue may be addressed. 
At redshifts $z>0.3$ we provide additional constraints on the X-ray
evolution of star-forming galaxies using the sub-sample of narrow
emission line radio sources with X-ray counterparts in the CDF-N
presented by Bauer et al. (2002).  These authors argue that the
properties of these radio galaxies are consistent with star-formation
activity dominating their X-ray emission. In the $j_X$ calculation we
consider sources in the redshift range $0.4<z<0.5$ with $S_{1.4}>50\mu 
\rm Jy$ (i.e. the $5\sigma$ radio completeness limit) and $f_X(\rm
0.5-8\,keV)>1\times10^{-16}\rm erg\,s^{-1}\,cm^{-2}$ (X-ray detection
limit). Only 4 sources in the Bauer et al. (2002) sample satisfy these
criteria and therefore the estimated $j_X$ plotted in Figure
\ref{fig_jx} suffers from large uncertainties. Despite the small
number statistics the $j_X$ of the CDF-N X-ray detected radio sources at
$z\approx0.5$ is elevated compared to both local spirals and
$z\approx0.3$ radio sources. The significance level of this
result is low but it is interesting that to the first approximation
the $j_X$ of X-ray {\it detected} radio  sources is in broad agreement
with the stacking analysis results presented here. 

Under the assumption that the faint radio population is indeed
dominated by star-formation activity, the observed increase in $j_X$
at least to $z\approx0.3$ suggests an  evolutionary rate of the form
$(1+z)^{3}$  broadly consistent with the Peak-M models of Ghosh \&
White (2001). This set of models is also in agreement with the
stacking analysis results of $z\approx0.5$ spirals  
of Brandt et al. (2001b).   

Assuming that the faint radio population is dominated by
star-formation activity (as the present study indicates) the increase
in $j_X$ with redshift is attributed to the evolution of the global
star-formation rate. Using the empirical relation $\rm
SFR=2.2\times10^{-40}\,\times\,L_X(\rm 0.5-2\,keV)\,\,M_{\odot}\,yr^{-1}$
(Ranalli et al. 2003) we estimate a SFR density of $0.029\pm0.007 \rm  
M_{\odot}\,yr^{-1}\,Mpc^{-3}$ for the $z=0.0-0.3$ redshift bin at a
median redshift of 0.240 in good agreement with previous studies
selecting star-forming galaxies at UV, optical or radio
wavelengths. This is demonstrated in Figure
\ref{fig_sfr}, showing the SFR density as a function of redshift. 


\section{Conclusions}\label{conclusions}
In this paper we apply stacking analysis to study the mean X-ray
properties of sub-mJy radio galaxies using a 50\,ks XMM-{\it
Newton} pointing overlapping  with a subregion of a deep and
homogeneous radio survey  reaching $\mu$Jy sensitivities (Phoenix Deep 
 Survey). Multiwavelength UV, optical and NIR 
photometric data are available for this field allowing photometric
redshifts and spectral types (i.e. ellipticals, spirals) to be
estimated for all radio sources brighter than $R=21.5$\,mag (total of
82). 

The subsample of $R<21.5$\,mag faint radio sources with spiral
galaxy SEDs (total of 34, after excluding AGN dominated sources) is
segregated into two redshift bins with a median of $z=0.240$ and 0.455
respectively. Stacking analysis is used to study the mean X-ray
properties of the sources in each redshift subsample. In the
0.5--2\,keV soft band a statistically significant signal is obtained
for both the low and the high-$z$  subsamples. Stacking of the hard
band counts yields a marginally significant signal ($2.6\sigma$) for
the $z=0.455$ subsample only.   

We argue that the AGN contamination of our
sample is minimal on the basis of the following arguments:  

\begin{enumerate}

\item Radio selection at $\mu$Jy flux densities has been shown to be
efficient in finding actively star-forming systems to
$z\approx1$. Only $\approx10-20$ per cent of the  faint radio population are
believed to be AGNs.   

\item Radio sources with optical counterparts exhibiting unresolved
optical light profile, likely to be 
distant QSOs, have been identified and excluded from the sample.  

\item Radio sources having X-ray counterparts with X-ray properties 
suggesting AGN activity have been excluded from the stacking
analysis. Similarly, faint radio sources with optical spectral
features (e.g. diagnostic emission line ratios, broad optical lines)  
typical of AGNs have also been excluded from the stacking sample. 

\item The mean X-ray properties of spectroscopically confirmed
star-forming  radio sources at $z<0.3$ are similar to those
obtained for the {\it whole} sample of $z<0.3$ radio sources with
spiral galaxy SEDs.

\item The mean X-ray and radio luminosities of the faint radio sources
studied here are consistent with the $L_X-L_{1.4}$ correlation of
local star-forming galaxies. 

\item The mean X-ray luminosity and X-ray--to--optical flux ratio of
the present faint radio sample are consistent with star-formation
rather than AGN activity.  

\end{enumerate}
 
Although the evidence above suggests that the observed X-ray emission
is dominated by star-formation activity we cannot exclude the
possibility of contamination of the present sample by a small number
of low-luminosity or heavily obscured  AGN. Our main conclusions are
summarised below:

\begin{itemize}

\item The mean $L_X/L_B$ ratio of the faint radio population studied
here  is elevated compared to optically  selected spirals in the
redshift range $0<z<1$. Differences in the $L_B$ distributions of the
samples can partly compensate for this effect. Elevated  
star-formation activity in the radio selected sample may also
be responsible for the increased $L_X/L_B$.

\item There is some evidence of increasing $L_X/L_B$ with increasing
redshift or $L_{1.4}$ for the sub-mJy galaxies. Although the
uncertainties hamper a secure interpretation this may suggest that
more energetic systems (in terms of   X-ray--to--optical flux ratio)
are also more powerful radio emitters.  

\item We find strong evidence for enhanced X-ray emissivity of
sub-mJy sources to $z\approx0.3$ compared to local H\,II galaxies and
$z\approx0.1$ spirals. This increase is consistent with X-ray
evolution of the form $\approx(1+z)^{3}$. If the present sample of
faint radio sources to $z\approx0.3$ is indeed dominated by
star-formation then this is the first direct evidence for evolution of
these systems at X-ray wavelengths. At $z>0.3$ incompleteness affects
our results and the present data cannot be used to
constrain the evolution in this regime. 

\item Using an empirical X-ray luminosity to SFR conversion factor we
estimate a global SFR density of  of $0.029\pm0.007 \rm
M_{\odot}\,yr^{-1}\,Mpc^{-3}$ at $z\approx0.3$. This is found to be in
fair agreement with previous studies.  
\end{itemize}

\section{Acknowledgments}
 We thank the anonymous referee for constructive comments.
 This work is jointly funded by the European Union  and the Greek
 Government  in the framework of the programme ``Promotion  of
 Excellence in Technological Development and Research'', project 
 ``X-ray Astrophysics with ESA's mission XMM''. AMH acknowledges
 support 
 provided by the National Aeronautics and Space Administration through
 Hubble Fellowship grant HST-HF-01140.01-A awarded by the Space 
 Telescope Science Institute.  JA gratefully acknowledges the support
 from the Science and Technology Foundation (FCT, Portugal) through
 the fellowship BPD-5535-2001 and the research grant
 POCTI-FNU-43805-2001.

\end{document}